\documentclass[12pt]{article}
\usepackage{amssymb}
\usepackage{bbm}
\usepackage{epsfig}
\usepackage{array}
\usepackage{float}
\usepackage{dsfont}
\usepackage{amstext}
\usepackage{psfrag}
\usepackage{a4}
\usepackage{a4wide}

\def\be{\begin{equation}}
\def\ee{\end{equation}}
\def\bea{\begin{eqnarray}}
\def\eea{\end{eqnarray}}

\begin{document}

\begin{center}
\baselineskip 20pt 
{\Large\bf Inflation in Supersymmetric SU(5)}

\vspace{1cm}

{\large 
S. Khalil$^{a,b}$, M. U. Rehman$^{c}$,
Q. Shafi$^{c}$ and E. A. Zaakouk$^{a}$
} 
\vspace{.5cm}

{\baselineskip 20pt \it
$^a$Ain Shams University, Faculty of Science, Cairo, 11566, Egypt \\
\vspace{2mm} 
$^b$Center for Theoretical Physics at the
British University in Egypt, \\
Sherouk City, Cairo 11837, Egypt \\ 
\vspace{2mm} 
$^c$Bartol Research Institute, Department of Physics and Astronomy, \\
University of Delaware, Newark, DE 19716, USA \\
}

\vspace{1cm}
\end{center}

\begin{abstract}
We analyze the adjoint field inflation in 
supersymmetric (SUSY) $SU(5)$ model.
In minimal SUSY $SU(5)$ hybrid inflation monopoles are 
produced at the end of inflation. We therefore
explore the non-minimal model of inflation based on
SUSY $SU(5)$, like shifted hybrid inflation, 
which provides a natural 
solution for the monopole problem. 
We find that the supergravity corrections with non-minimal 
K\"ahler potential are crucial 
to realize the central value of the scalar 
spectral index $n_{s}\simeq 0.96$ consistent with 
the seven year WMAP data. The tensor to scalar ratio $r$
is quite small, taking on values $r \lesssim 10^{-5}$. 
Due to $R$-symmetry massless $SU(3)$ octet and $SU(2)$
triplet supermultiplets are present and could spoil
gauge coupling unification. To keep gauge coupling 
unification intact, light vector-like particles are added 
which are expected to be observed at LHC.
\end{abstract}

%
\section{\large{\bf Introduction}}%
Inflation is one of the most motivated scenarios for the early
universe, which is consistent with the recent cosmological
observations on the cosmic microwave background radiation and the
large-scale structure in the universe. In order to construct a
consistent model of inflation, an extension of the Standard Model
(SM) is required. The supersymmetric (SUSY) grand unified theory
(GUT) models provide a natural framework for hybrid inflation 
\cite{Linde:1993cn}-\cite{Lyth:1998xn}.
On the other hand, supersymmetric $SU(5)$ is the 
simplest extension of the SM which may realize 
hybrid inflation through the adjoint scalar field, 
which is responsible for breaking the $SU(5)$ gauge symmetry into 
SM gauge group $G_{SM}\equiv SU(3)_c\times SU(2)_L\times U(1)_Y$.
However, in the standard minimal version of SUSY hybrid
inflation the gauge symmetry is broken
at the end of inflation, and topological 
defects are copiously formed.
To overcome this problem, we consider the leading
order non-renormalizable term in the superpotential of 
SUSY $SU(5)$ hybrid inflation. This class of inflationary models 
is known as shifted hybrid inflation, which have been introduced in
Ref.~\cite{Jeannerot:2000sv} in the context of a supersymmetric 
Pati-Salam model. 
The inclusion of the non-renormalizable term introduces 
a non-trivial flat direction along which $SU(5)$ gauge
symmetry is spontaneously broken through the
appropriate Higgs fields acquiring non-zero 
vacuum expectation values (vevs).
This direction then can be used as an inflationary trajectory 
with one-loop radiative corrections providing the necessary 
slope for slow-roll inflation.
However, since $SU(5)$ is broken during inflation, one finds 
that for a certain range of parameters, the system always passes 
through the above mentioned inflationary trajectory before falling 
into the SUSY vacuum.
Therefore, the magnetic monopole problem is solved for all initial
conditions.

The scalar spectral index $n_{s}$ in shifted hybrid inflation, 
driven solely by radiative corrections, 
is typically of order $0.98$ with 
the number of e-foldings $N_0 \simeq 53$ and lies outside the
seven year Wilkinson Microwave Anisotropy Probe (WMAP7) 
1-$\sigma$ bounds \cite{Komatsu:2010fb}. 
Including supergravity (sugra) corrections with 
minimal K\"ahler potential 
further enhances the scalar spectral index to 
exceed unity and a blue-tilted scalar spectral index
is obtained as emphasized in 
Refs.~\cite{Senoguz:2003zw,Panagiotakopoulos:1997qd,Linde:1997sj,Kawasaki:2003zv}.
The non-minimal K\"ahler potential plays a crucial role in reducing 
$n_{s}$ and making it compatible with the WMAP7 data as shown in 
Refs.~\cite{BasteroGil:2006cm,urRehman:2006hu,Shafi:2010jr}.
In SUSY $SU(5)$ shifted hybrid inflation we show that the 
central value of the scalar spectral index $n_s \simeq 0.96$
can be realized in the presence of a small negative mass 
term generated from a non-minimal K\"ahler potential.

A salient feature of above mentioned $SU(5)$ inflationary
model is that the octet and triplet components of the adjoint
scalar field remain massless below GUT scale after the
breaking of $SU(5)$ symmetry. The masses of these particles 
are related to the scale of $R$-symmetry breaking. 
We discuss the effect of these particles on the gauge coupling 
unification along with the extra vector-like particles 
which should be added to restore unification. 
If $R$-symmetry is broken at TeV scale, 
these particles can be observed at the Large Hadron Collider 
(LHC) with clear signatures.

The paper is organized as follows. 
In section 2 we discuss $SU(5)$ shifted hybrid inflation.
Here we explore conditions which can lead inflaton 
to the shifted minimum. We also include sugra
corrections with the minimal and non-minimal K\"ahler potential. 
In section 3, we discuss the impact 
of octet and triplet, which remain massless after $SU(5)$ 
symmetry breaking, on gauge coupling unification and show how to 
restore unification by introducing extra vectorlike particles 
at TeV scale. Finally our conclusions are given in section 4.

%
%
\section{\large{\bf Shifted hybrid $SU(5)$ inflation}}%
In SUSY $SU(5)$, the matter fields are assigned to the
$1$, $\bar{5}$ and $10$ dimensional representations, while the Higgs
fields belong to the adjoint scalar $\Phi\,( \equiv 24_{H})$ and
fundamental scalars: $H\,(\equiv 5_{h})$ and $\bar{H}\,(\equiv \bar{5}_{h})$.
The $R$-symmetric\footnote{The SUSY
$SU(5)$ inflation with $R$-symmetry violating terms has been
considered in Ref.~\cite{Covi:1997my}.} $SU(5)$ superpotential, 
with the leading order non-renormalizable term, is given by
\bea%
W= S\left[\kappa M^2-\kappa Tr(\Phi^{2})-\frac{\beta}{M_*}
Tr(\Phi^3)\right]+\gamma \bar{H}\Phi H+\delta \bar{H} H  \nonumber \\
+ y_{ij}^{(u)}\,10_i\,10_j\,H + y_{ij}^{(d,e)}\,10_i\,\bar{5}_j\,\bar{H}
+y_{ij}^{(\nu)}\,1_i\,\bar{5}_j\,H + m_{\nu_{ij}}\,1_i\,1_j ,
\eea %
where $S$ is a gauge singlet superfield, $M_*$ is some 
suitable cutoff scale, $m_{\nu_{ij}}$
is the neutrino mass matrix and $y_{ij}^{(u)}$, 
$ y_{ij}^{(d,e)}$, $y_{ij}^{(\nu)}$ are the Yukawa 
couplings for quarks and leptons. Only the terms 
linear in $S$ are relevant for inflation,
and their role in realizing successful inflation
will be discussed in detail. The other two terms
in the first line in Eq.~(1) are involved in the 
solution of the doublet-triplet problem. Fine tuning
is required to implement doublet-triplet splitting
and thereby adequately suppress proton decay.
The second line contains terms that generate masses 
for quarks and leptons. The $R$-charge assignments
of the various superfields are as follows:
\be %
(R_S, R_{\Phi}, R_{H}, R_{\bar{H}}, R_{10}, R_{\bar{5}}, R_{1}) 
= \left(1, 0, \frac{2}{5}, \frac{3}{5}, \frac{3}{10}, \frac{1}{10}
, \frac{1}{2}\right). %
\ee %

In component form, the above superpotential takes the following form%
\be%
W \supset S\left[\kappa M^2-\kappa\frac{1}{2}
\sum_{i}\phi_{i}^{2}-\frac{\beta}{4M_*}d_{ijk}\phi_{i}\phi_{j}
\phi_{k}\right]+\gamma T_{a b}^{i}\phi^{i}\bar{H_{a}}H_{b}+\delta
\bar{H_{a}}H_{a},%
\label{superpot-shift}%
\ee %
where we express the scalar field $\Phi$ in the $SU(5)$ adjoint
basis $\Phi= \phi_i T^i$ with $Tr(T_i T_j)=\delta_{ij}/2$ and 
$d_{ijk}=2 Tr(T_i\{T_j,T_k\})$. Here the indices $i$, $j$, $k$
run from 1 to 24 whereas the indices $a$, $b$ run from 1 to 5.
Moreover, the repeated indices are summed over.
The scalar potential obtained from this superpotential 
is given by %
\begin{eqnarray}%
V &\supset& \kappa^2 \left| M^2-\frac{1}{2}
\sum_{i}\phi_{i}^{2}-\frac{\beta}{4 \kappa
M_*}d_{ijk}\phi_{i}\phi_{j}
\phi_{k} \right|^{2}+\sum_{i}\left|\kappa S \phi_{i}+\frac{3
\beta}{4 M_*}d_{ijk} S \phi_{j} \phi_{k}
- \gamma T_{a
b}^{i}\bar{H_{a}}H_{b}\right|^{2}\nonumber\\&&
 + \sum_{b}\left|\gamma
T_{a b}^{i}\phi^{i}\bar{H_{a}}+\delta
\bar{H_{b}}\right|^{2}+\sum_{b}\left|\gamma T_{a
b}^{i}\phi^{i}H_{a}+\delta
H_{b}\right|^{2} + D\text{-terms}.%
\label{scalarpot-shift}
\end{eqnarray}%
Note that the scalar components of the superfields are 
denoted by the same symbols as the corresponding superfields.
Vanishing of the $D$-terms is achieved with $\vert \bar{H}_a \vert =
\vert H_a \vert$ and $\phi_i =\phi_i^*$. We restrict ourselves to
this $D$-flat direction and use an appropriate $R$ transformation 
to rotate $S$ complex field to the real axis, $S = \sigma/\sqrt{2}$, 
where $\sigma$ is a real scalar field. The supersymmetric
global minimum of the above potential lies at%
\be %
\sigma^0 =H_{a}^{0}=\bar{H_{a}^{0}}=0, %
\ee %
with $\phi_i^0$ satisfying the following equation: %
\be%
\sum_{i=1}^{24}(\phi_{i}^{0})^{2} + \frac{\beta}{2 \kappa M_*}
d_{ijk}
\phi^0_i \phi^0_j \phi^0_k =2M^{2}.
\ee%
The superscript `0' denotes the field value at its
global minimum.
Using $SU(5)$ transformation, one can express the vev matrix 
$\Phi^0$ in diagonal form with $\phi_i^0 \neq 0$ for 
the diagonal generators only. Now in order to break $SU(5)$ gauge 
symmetry into the SM gauge group $G_{SM}$, the vevs of all 
$\phi_i^0$ components must vanish except the one which is
invariant under $G_{SM}$. 
Therefore, we choose $\phi_{24}^0$ to have a non-vanishing vev:
$\phi_{24}^0 = v/\sqrt{2}$ where $v$ satisfies the following equation:%
\be%
v^2 - \frac{\beta}{2\sqrt{30} M_* \kappa} v^3 = 4 M^2.%
\label{vev}
\ee%
Here, we have used the fact that $d_{24,24,24}=-1/\sqrt{15}$.
We can rewrite the scalar potential in
Eq.~(\ref{scalarpot-shift}) in terms of the dimensionless
variables%
\be %
y = \frac{\phi_{24}/M}{\sqrt{2}}~, ~~~~~~~~~~~~ 
z = \frac{\sigma/M}{\sqrt{2}} %
\ee %
as, 
\be %
\tilde{V} = \frac{V}{\kappa^2 M^4} = 
\left(1-y^2 + \xi y^3 \right)^2 +  2 z^2 y^2
\left(1- 3\xi y/2\right)^2~,%
\ee %
where $\xi = \beta M/\sqrt{30}\kappa M_*$. Thus, for a constant
value of $z$, this potential has the following three extrema: %
\bea %
y_1 &=& 0,\\
y_2 &=& \frac{2}{3 \xi},\\
y_3 &=& \frac{1}{3 \xi} - \frac{2^{1/3} (-1+ 9 \xi^2 z^2)}{3\xi
\left(2-27\xi^2 +\sqrt{(2-27\xi^2)^2 +4(-1+9
\xi^2 z^2)^3}\right)^{1/3}}\nonumber\\
&+& \frac{\left(2-27\xi^2 +\sqrt{(2-27\xi^2)^2 +4(-1+9
\xi^2 z^2)^3}\right)^{1/3}}{3 2^{1/3} \xi}.%
\eea%

\begin{figure}[t]
\centering \includegraphics[width=10cm]{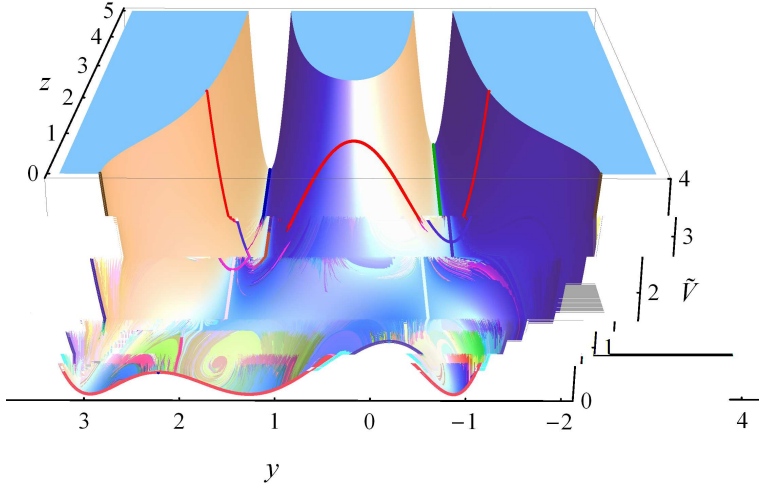}
\caption{The dimensionless potential 
$\tilde{V}=V/\kappa^2M^4$ versus $y$ and $z$ for $\xi= 0.3$.
The red curve at $z = 2$ clearly shows one maximum at $y \sim 1$
and two minima at $y_1 = 0$ (standard trajectory) and 
$y_2 = 2/3\xi = 0.22$ (shifted trajectory). These minima become 
maxima near $z = 0$ as are shown by the brown curve.}
\label{fig1}
\end{figure}

The first two extrema at $y_1$ and $y_2$ are independent of $z$ (or $|S|$)
and correspond to the `standard' and the `shifted' inflationary trajectories.
The extremum at $y_1$ is a local minimum (maximum) 
for $z > 1 \, (z < 1)$, while the shifted extremum at
$y_2$ is a local minimum (maximum) for 
$z^2 > 4/27\xi^2 -1 \, (z^2 < 4/27\xi^2 -1) $. 
These trajectories are shown in Fig.~\ref{fig1} where we have
plotted the dimensionless potential $\tilde{V}$ as a function 
of $z$ and $y$ for a typical value of $\xi=0.3$.
The potential at $y_2$ is $\tilde{V}_2=(1-4/27\xi^2)^2$, 
which is lower than the potential $\tilde{V}_1=1$ at 
$y_1$ for $\xi > \sqrt{2/27}\simeq 0.27$. 
Inflation takes place when the system is trapped along the
$y_2$ minimum. Also, we restrict ourselves to $\xi <
\sqrt{4/27}\simeq 0.38 $, so that the inflationary trajectory at
$y_2$ can be realized before $z$ reaches zero. Therefore, the
interesting region for the parameter $\xi$ in our analysis is
given by $0.27 < \xi < 0.38$. Moreover, the $SU(5)$ gauge symmetry
is always broken during the inflationary trajectory and hence 
no magnetic monopoles are produced at the end of inflation.


Along the shifted trajectory SUSY is broken due to the presence
of a non-zero vacuum energy density $\kappa^2M^4(1-4/27\xi^2)^2$. 
This in turn generates the radiative corrections which can lift the
flatness of the $y_2$ trajectory while providing the necessary slope
for driving inflation. In order to calculate the one-loop radiative 
correction at $y_2$ we need to compute the mass spectrum of the model 
along this path where both $SU(5)$ gauge symmetry and SUSY are broken.

During inflation, the field $\Phi$ acquires a vev 
in the $\phi_{24}$ direction which breaks the 
$SU(5)$ gauge symmetry down to SM gauge group $G_{SM}$. 
Perturbing around this vacuum $v_2 = 2M y_2$ and replacing 
$\phi_{24} \rightarrow \left[ v_2 
+ Re(\phi_{24}) + i \, Im(\phi_{24}) \right]/\sqrt{2}$, 
the potential in Eq.~(\ref{scalarpot-shift}) 
yields the following masses for real scalar fields, 
$Re(\phi_{24})$ and $Im(\phi_{24})$:%
\be%
m^{2}_{24_{\pm}}=\pm
\kappa^{2}M^{2}(\frac{4}{27\xi^{2}}-1)+\kappa^{2}|S|^{2}.%
\ee%
The superpotential in Eq.~(\ref{superpot-shift})
generates a Weyl fermion with mass-squared: 
\be
m_{24}^2=\kappa^{2}|S|^{2}.
\ee
Similarly, we obtain the following masses for the real scalar
fields $Re(\phi_{i})$ and $Im(\phi_{i})$ with $i=1,...,8,21,22,23$:
\be %
m^{2}_{i_{\pm}}=\pm 5 \kappa^{2}M^{2}(\frac{4}{27\xi^{2}}-1)+25
\kappa^{2}|S|^{2},%
\ee%
and from the following terms of the superpotential%
\be%
\delta W = \kappa S \left( M^2 - \frac{1}{2} \phi_{i}^{2} 
- \frac{3\beta}{4 \kappa M_*}
d_{i\,i\,24} \phi_{i}^2 \phi_{24}\right),%
\ee%
the $11$ weyl fermions $\psi_{i}$, $i=1,...,8,21,22,23$, acquire 
a universal mass-squared:
\be%
m^{2}_i= 25 \kappa^{2}|S|^{2}.%
 \ee%
It is worth noting that the SUSY breaking along the inflationary
trajectory, which is due to the non zero vacuum energy 
$\kappa^{2}M^{4}(1-\frac{4}{27\xi^{2}})^{2}$, generates a mass splitting
in $\phi_{24}$ supermultiplet and in $\phi_{i}$,
$i=1,...,8,21,22,23$, supermultiplets. This is the only place where
mass splitting between fermions and bosons appears.

The $D$-term contribution to the masses of scalar fields $\phi_{i}$,
$i=9,...,20$, is obtained from the following term:%
\be%
g ^{2} \left( f^{ijk} \phi_{j} \phi^{\dag}_k \right)
\left( f^{ilm} \phi_{l} \phi^{\dag}_m\right),%
\ee%
where $g$ is the $SU(5)$ gauge coupling. As an example, we
obtain the mass of $\phi_{10}$ field as follows:%
\be%
\frac{1}{2}\,g ^{2} \left( f^{9\,24\,10} v_2 \phi_{10}+f^{9\,10\,24} v_2
\phi^{\dag}_{10} \right)^{2}, %
\ee%
which leads to mass-squared $\frac{25}{30} g^{2} v_2^2$ using
$f^{9\,10\,24} = - f^{9\,24\,10} = \frac{1}{2}\sqrt{\frac{5}{3}}$
and $f^{i\,10\,24} = 0 $ for $i = 10,...,20$. 
Thus, the D term contributes with a universal 
mass-squared $\frac{25}{30} g^{2} v_2^2$ for 
the above mentioned $12$ real scalar fields. 
The mixing between chiral fermions $\psi_{i}$, $i=9,....,20$,
and the gauginos $\lambda_{i}$, $i=9,....,20$, gives rise to
Dirac mass term:%
\be%
i\sqrt{2} g f^{ijk} \left( \phi^{\dag}_k \lambda_{i}
\psi_{j} + \bar{\psi}_{k} \bar{\lambda}_{i} \phi_{j}\right).%
\ee%
This leads to $12$ Dirac fermions with mass-squared 
$\frac{25}{30}g^{2}v_2^2$. Finally, the gauge bosons
$A_{\mu}^{i}$, $i=9,...,20$, acquire masses from the following
term: 
\be%
g^{2} f^{ijk} f^{ilm} A_{\mu}^{j}A^{l \mu} \phi^{\dag}_k
\phi_{m}.
\ee%
This generates a universal mass-squared $\frac{25}{30} g^{2}v_2^2$ for all
$12$ gauge bosons.

In Table 1, we summarize the results of the mass spectrum of the
model along the shifted inflationary trajectory. 
As noted above, the mass splitting only occurs in 
$\phi_{24}$ and $\phi_i$, $i=1,...,8,21,22,23$, supermultiplets
which contain 12 Majorana fermions with two degrees of freedom 
and 24 real scalars, whereas there is no mass splitting
in $\phi_i$, $i=9,...,20$, supermultiplets which consist of
12 massive Dirac fermions with four degrees of freedom, 
12 massive gauge bosons with 3 degrees of freedom, 
and 12 real scalars. It can be readily checked that
all these supermultiplets satisfy the supertrace rule $StrM^{2}=0$.
\begin{table}[t]
\centering {
\begin{tabular}{c||c}
\hline\hline {\bf Fields} ~~~~~~& ~~~~~~~~ {\bf Squared Masses} \\
\hline 2 real scalars ~~~~~~~& ~~~~~~~~$\kappa^2 \vert S\vert^2 \pm \kappa^2 M^2 (\frac{4}{27 \xi^2}-1)$ \\
\hline 1 Majorana fermion ~~~~~~~& ~~~~~~~~$\kappa^2 \vert S\vert^2 $ \\
\hline 22 real scalars ~~~~~~~& ~~~~~~~~$25 \kappa^2 \vert S\vert^2 \pm 5\kappa^2 M^2 (\frac{4}{27 \xi^2}-1)$ \\
\hline 11 Majorana fermions ~~~~~~~& ~~~~~~~~$25 \kappa^2 \vert S\vert^2 $ \\
\hline 12 real scalars ~~~~~~~& ~~~~~~~~$ \frac{25}{30} g^2 v_2^2 $ \\
\hline 12 Dirac fermions ~~~~~~~& ~~~~~~~~$\frac{25}{30} g^2 v_2^2 $ \\
\hline 12 gauge bosons ~~~~~~~& ~~~~~~~~$\frac{25}{30} g^2 v_2^2 $ \\
\hline\hline
\end{tabular}%
} \caption{The mass spectrum of the shifted hybrid $SU(5)$ model as the
system moves along the inflationary trajectory 
$y_2$ ($v_2 = 2\,M\,y_2 = \frac{4\,M}{3\,\xi}$).} \label{spectrum}
\end{table}

The inflationary effective potential with
1-loop radiative correction is given by %
\bea %
V_{\rm 1 loop} \!\!&\!=\!&\!\! \kappa^2 M_{\xi}^4 \left(1+
\frac{\kappa^2}{16 \pi^2} \left[ F(M_{\xi}^2,x^2)
+11\times 25\,F(5 M_{\xi}^2,5\,x^2) \right]\right),%
\label{Vloop}
\eea%
where
\be %
F(M_{\xi}^2,x^2) = 
\frac{1}{4}\left( \left( x^4+1\right) \ln \frac{\left( x^4-1\right) 
}{x^4}+2x^2\ln \frac{x^2+1}{x^2-1}+2\ln \frac{\kappa ^{2}M_{\xi}^{2}x^2}{%
Q^{2}}-3\right),
\ee%
$x=|S|/ M_{\xi}$, $M_{\xi}^2 = M^2 (4/27 \xi^2 -1)$ and
$Q$ is the renormalization scale.
The spectrum of the model at the $SU(5)$ breaking SUSY
minimum is given by the massless octet $\phi_i$,
$i=1,..,8$, and triplet $\phi_k$, $k=21,22,23$ scalars/fermions, 
while the fields $\phi_j$, $j=9,...,20$, acquire mass-squared of 
order $g^2 v_2^2$. Finally $\phi_{24}$ and $S$ fields acquire 
masses of order $\kappa M$. 
As will be shown later, these octets and triplets spoil the gauge coupling
unification and we, therefore, need to add some vector-like matter 
to preserve unification. Before discussing unification we consider
the contribution from sugra corrections.
\subsection{\large{\bf Sugra corrections and non-minimal K\"ahler
potential}}
We take the following general form of K\"ahler potential:%
\bea %
K &=& \vert S \vert^2 + Tr \vert \Phi \vert^2 
+ \vert H \vert^2 + \vert \bar{H}\vert^2  \nonumber \\
&&+ \kappa_{S\Phi} \frac{\vert S\vert^2 \, Tr \vert \Phi \vert^2}{m_P^2}
+ \kappa_{S H} \frac{\vert S \vert^2 \vert H \vert^2}{m_P^2}
+ \kappa_{S \bar{H}} \frac{\vert S \vert^2 \vert \bar{H} \vert^2}{m_P^2} \nonumber \\
&& + \kappa_{H \Phi} \frac{\vert H \vert^2 \, Tr \vert \Phi \vert^2}{m_P^2} 
+ \kappa_{\bar{H} \Phi} \frac{\vert \bar{H} \vert^2 \, Tr \vert \Phi \vert^2}{m_P^2} 
+ \kappa_{H \bar{H}} \frac{\vert H \vert^2 \vert \bar{H} \vert^2}{m_P^2} \nonumber \\
&& + \kappa_S \frac{\vert S\vert^4}{4 m_P^2} 
+ \kappa_{\Phi} \frac{ (Tr \vert \Phi \vert^2)^2}{4 m_P^2} 
+ \kappa_{H} \frac{ \vert H \vert^4}{4 m_P^2} 
+ \kappa_{\bar{H}} \frac{ \vert \bar{H} \vert^4}{4 m_P^2} \nonumber \\
&& + \kappa_{SS} \frac{\vert S\vert^6}{6 m_P^4} 
+ \kappa_{\Phi \Phi} \frac{ (Tr \vert \Phi \vert^2)^3}{6 m_P^4} 
+ \kappa_{H H} \frac{ \vert H \vert^6}{6 m_P^4} 
+ \kappa_{\bar{H} \bar{H}} \frac{ \vert \bar{H} \vert^6}{6 m_P^4}
+ \cdots , %
\label{K}
\eea %
where $m_P \simeq 2.4 \times 10^{18}$ GeV is the reduced Planck mass.
Additionally, for the sake of simplicity, the contribution 
of many other terms e.g., of the form 
\be 
c_2\,[Tr(\Phi^2) + h.c.] + c_3\,[Tr(\Phi^3)/m_P + h.c.] + \cdots
\ee
is assumed to be zero. Alternatively we can 
effectively absorb these extra contributions coming 
from the $\Phi$ superfield into various couplings of the 
above K\"ahler potential as only the $|S|$ field plays an
active role during inflation.
The sugra scalar potential is given by
\begin{equation}
V_{F}=e^{K/m_P^{2}}\left(
K_{i\bar{j}}^{-1}D_{z_{i}}WD_{z^{*}_j}W^{*}-3 m_P^{-2}\left| W\right| ^{2}\right),
\label{VF}
\end{equation}
with $z_{i}$ being the bosonic components of the superfields $z%
_{i}\in \{S,\Phi,H,\bar{H},\cdots\}$ and where we have defined
\be
D_{z_{i}}W \equiv \frac{\partial W}{\partial z_{i}}+m_P^{-2}\frac{%
\partial K}{\partial z_{i}}W , \,\,\,
K_{i\bar{j}} \equiv \frac{\partial ^{2}K}{\partial z_{i}\partial z_{j}^{*}},
\ee
and $D_{z_{i}^{*}}W^{*}=\left( D_{z_{i}}W\right)^{*}.$ 
Now in the inflationary trajectory with the D-flat direction 
($\phi_i = \phi_i^*$, $|\bar{H}_a| = |H_a|$)
and using Eqs. (\ref{superpot-shift}), (\ref{Vloop}), (\ref{K}), and (\ref{VF}),
we obtain the following form of the full potential:
\bea
V &=& \kappa ^{2}M_{\xi}^{4}\left( 1 +
\frac{\kappa^2}{16 \pi^2} \left[ F(M_{\xi}^2,x^2)
+11\times 25\,F(5M_{\xi}^2,5\,x^2) \right] \right. \nonumber \\
&& + \left(\frac{4(1-\kappa_{S\Phi})}{9\,(4/27 - \xi^2)} -\kappa_S\,x^2 \right)
\left( \frac{M_{\xi}}{m_P}\right)^{2} \nonumber \\
&& \left. + \left( \frac{4((1-2\kappa_{S\Phi})^2+1+\kappa_{\Phi})}{81\,(4/27 - \xi^2)^2}
 \right. \right. \nonumber \\
&& \left. \left. 
+\frac{4((1-\kappa_{S\Phi})^2-\kappa_{S}(1-2\kappa_{S\Phi}))x^2}{9\,(4/27 - \xi^2)}
+\frac{\gamma _{S}\,x^4}{2}\right) 
\left( \frac{M_{\xi}}{m_P}\right)^{4} +\cdots \right) +\cdots\,, 
\label{VT}
\eea
where $x = |S|/ M_{\xi}$, $M_{\xi}^2 = M^2 (4/27 \xi^2 -1)$ and
$\gamma _{S}=1-\frac{7\kappa _{S}}{2}+2\kappa _{S}^{2}-3\kappa _{SS}$.
We restrict ourselves to $\kappa \gtrsim 10^{-3}$ and 
neglect contribution of soft SUSY breaking terms 
assuming soft masses of order 1~TeV 
\cite{Senoguz:2004vu,Rehman:2009nq,Rehman:2009yj}.

Before proceeding further, let us consider
the possible mass contribution from the non-minimal 
terms of the K\"ahler potential. In particular, as we will see, 
the $SU(2)$ triplet and $SU(3)$ octet multiplets remain 
massless as a consequence of both $R$ and $SU(5)$ gauge symmetries.
To see this explicitly, consider the following general form of 
the fermionic mass matrix:
\be
({\cal M}_F)_{ij} = e^{K/2} \left( W_{ij} + K_{ij}W + K_iW_j + 
K_j W_i + K_i K_j W - K^{k\overline{l}}K_{ij\overline{l}} D_k W \right).
\ee
Since, in the SUSY minimum $W$ and $W_i$ essentially vanish due to 
R-symmetry, the triplet and octet multiplets therefore do not acquire 
masses from the above contribution. 
The other possible contribution to fermionic masses come
through the mixing of chiral fermions and gauginos, 
\be 
i\,g\,\sqrt{2}\,f^{i\,j\,k}\,\left(\phi_k\,\psi_{\bar{l}}\,K_{{\bar{l}}j}\lambda_i 
+ \bar{\psi}_{\bar{l}}\,K_{{\bar{l}}k}\,\bar{\lambda}_i\,\phi_j \right).
\ee
Due to $SU(5)$ gauge symmetry, these terms also 
vanish for the triplet and octet multiplets. 
(Note that $f^{i\,j\,24} = 0$ for the triplet and octet states).
Therefore, both the fermionic and bosonic masses of these multiplets
vanish as a consequence of $R$-symmetry and SUSY $SU(5)$.
However, these multiplets can acquire TeV scale masses as is discussed 
in more detail in the third section. 

On the other hand, in the shifted trajectory case 
a non-zero mass contribution is expected from the 
nonminimal terms of the K\"ahler potential.
But these contributions are expected to have
a negligible effect on the inflationary predictions as they
appear inside the logarithmic functions of radiative 
corrections. Therefore, in numerical calculations
we can safely ignore these contributions.

Before starting our discussion of this model it is useful 
to recall here the basic results of the
slow roll assumption. The inflationary slow-roll parameters are given by
\be
\epsilon = \frac{m_P^2}{4\,M_{\xi}^2}\left( \frac{\partial_x V}{V} \right)^2, 
\,\,\, \eta = \frac{m_P^2}{2\,M_{\xi}^2} \left( \frac{\partial_x^2 V}{V} \right).
\ee
In the slow-roll approximation (i.e. $\epsilon$, $|\eta| \ll 1$), 
the scalar spectral index $n_s$ and the tensor to scalar ratio $r$ 
are given (to leading order) by
\be
n_s \simeq 1 + 2\,\eta -6\,\epsilon, 
\,\,\, r \simeq 16\,\epsilon. 
\ee
The number of e-folds during inflation
$l_0 = 2\,\pi/k_0$ has crossed the horizon
is given by
\begin{equation}
N_0 = 2\left( \frac{M_{\xi}}{m_P}\right) ^{2}\int_{x_e}^{x_{0}}\left( \frac{V}{%
\partial _{x}V}\right) dx,
\end{equation}
where $|S_0|=x_0 \, M_{\xi}$ is the field value at 
the comoving scale $l_0$, and $x_e$ denotes the value of field at the 
end of inflation, defined by $|\eta(x_e)| = 1$ (or $x_e = 1$).
During inflation, this scale exits the horizon
at approximately
\begin{equation}
N_{0}=53+\frac{1}{3}\ln \left( \frac{T_{r}}{10^{9}~{\rm GeV}}\right) +\frac{2}{3}%
\ln \left( \frac{V^{1/4}(x_0)}{10^{15}~{\rm GeV}}\right),
\end{equation}
where $T_{r}$ is the reheat temperature and
for a numerical work we will set $T_r = 10^9$ GeV. This
could easily be reduced to lower values if the gravitino 
problem\footnote{For a recent discussion on the gravitino 
overproduction problem in hybrid inflation see 
Ref.~\cite{Nakayama:2010xf}.} is regarded to be an issue. 
The subscript `0' denotes the comoving
scale corresponding to $k_{0}=0.002$ Mpc$^{-1}$.
The amplitude of the curvature perturbation is given by
\begin{equation} \label{perturb}
\Delta_{\mathcal{R}}^2 = \frac{1}{24\,\pi^2} 
\left. \left( \frac{V/m_P^4}{\epsilon}\right)\right|_{k = k_0},
\end{equation}
where $\Delta_{\mathcal{R}}^2 = (2.43\pm 0.11)\times 10^{-9}$ is 
the WMAP7 normalization at $k_0 = 0.002\, \rm{Mpc}^{-1}$ 
\cite{Komatsu:2010fb}. 
Note that, for added precision, we include in our calculations 
the first order corrections \cite{NeferSenoguz:2008nn} 
in the slow-roll expansion 
for the quantities $n_s$, $r$, and $\Delta_{\mathcal{R}}$.

\begin{figure}[t]
\begin{center}
\epsfig{file=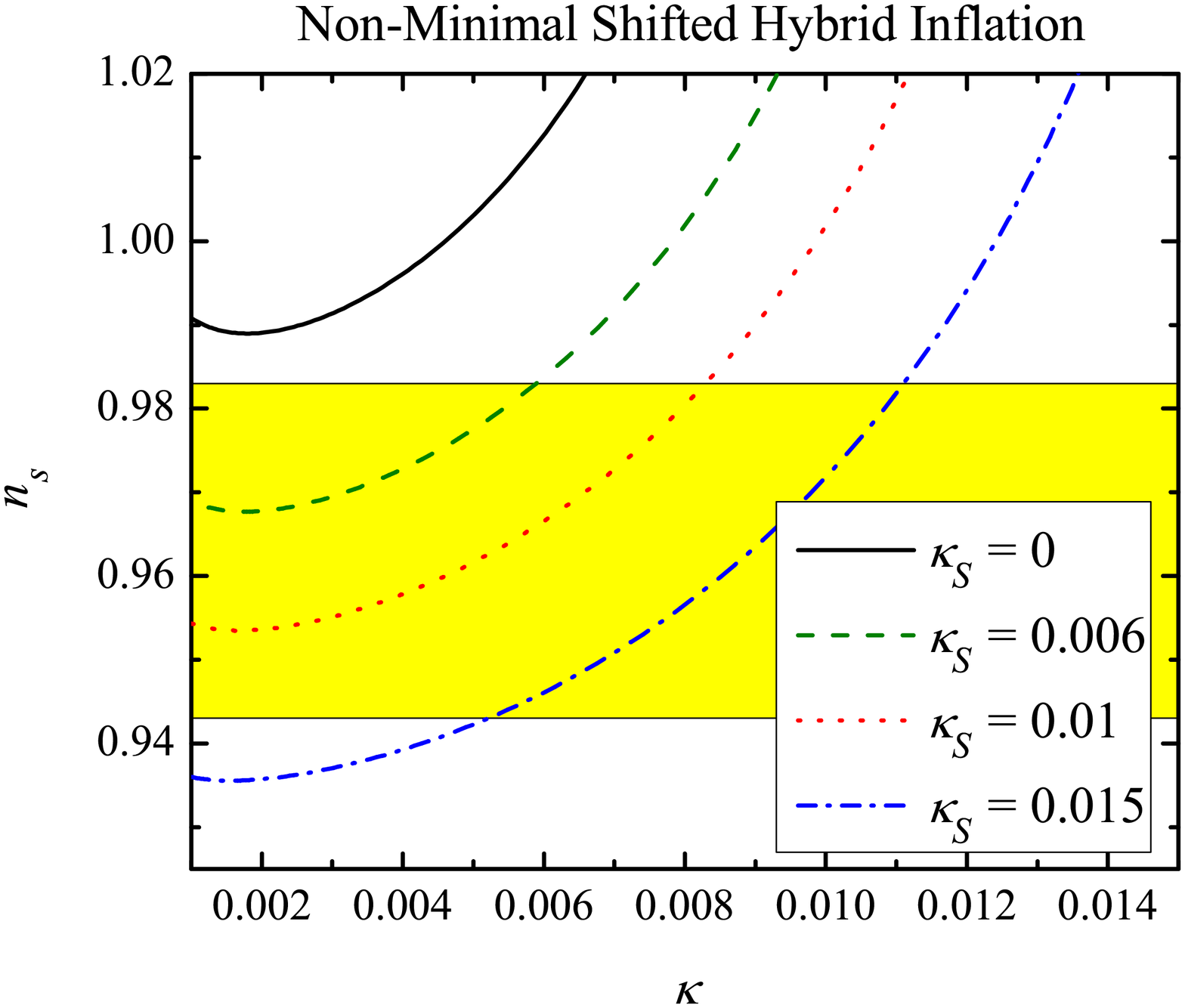, width=8cm}
\epsfig{file=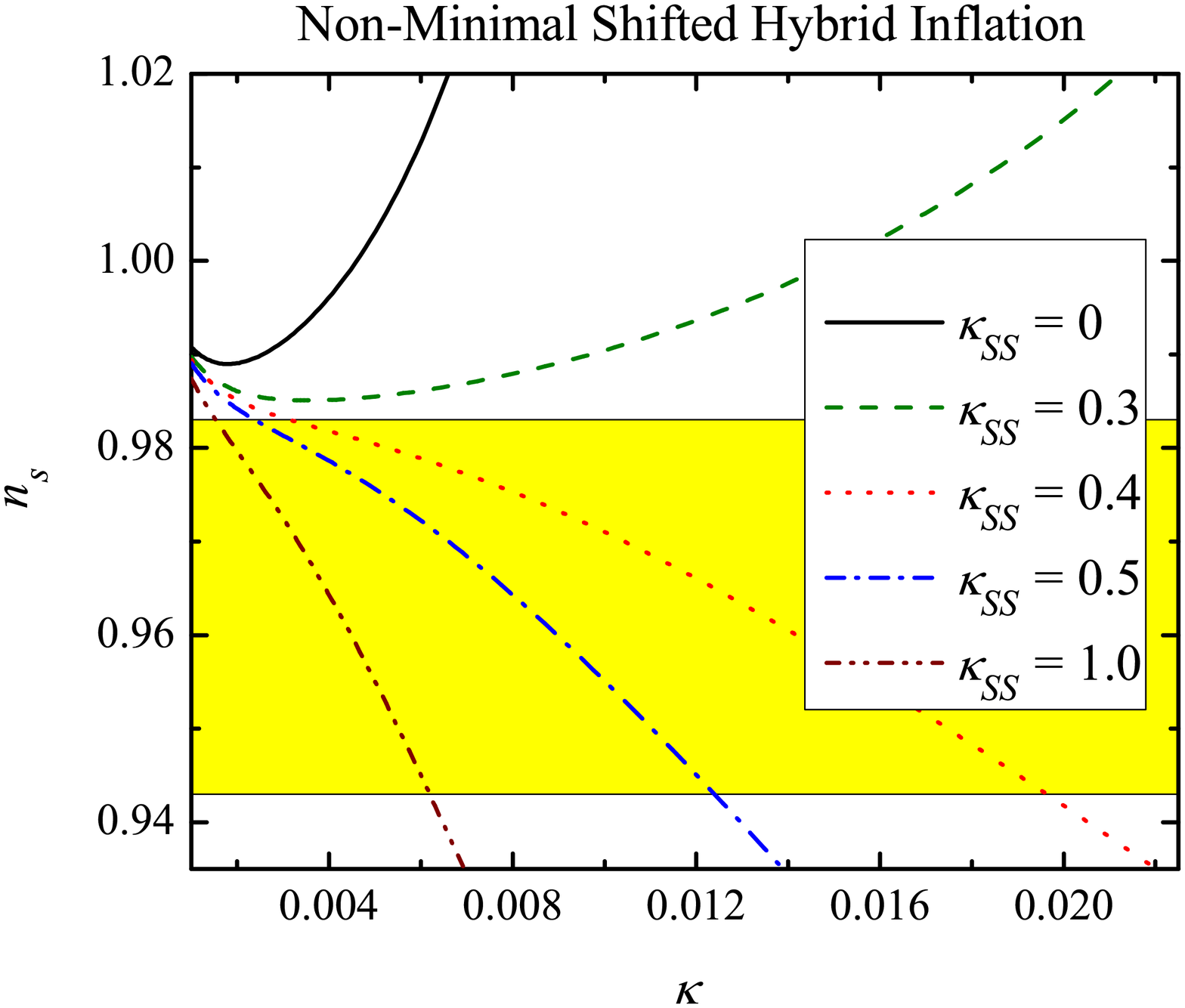, width=8cm}
\end{center}
\caption{$n_s$ vs $\kappa$
for shifted hybrid inflation with $\xi = 0.3$ and $T_r = 10^{9}$ GeV.
The WMAP7 1-$\sigma$ (68\% confidence level) bounds are
shown in yellow.}
\label{fig2}
\end{figure}
 
Including sugra correction, in general, introduces a mass squared 
term of order $H^2$, where $H \simeq \sqrt{V/3\,m_P^2}$ is the Hubble parameter.
This in turn makes the slow parameter $\eta \sim 1$ and spoils
inflation. This is the well known $\eta$ problem. However, in
the case of supersymmetric shifted hybrid inflation 
with minimal K\"ahler potential 
this problem is naturally resolved due to
a cancellation between the mass squared terms of the exponential
factor and the other part of the potential in Eq.~(\ref{VF}). 
The linear dependence of $W$ in $S$ due to $R$-symmetry
guarantees this cancellation to all orders 
\cite{Copeland:1994vg,Stewart:1994ts}. 
With non-minimal K\"ahler potential the mass 
squared term can be approximated as
\begin{equation}
\sim \left( \kappa_S + 
\frac{4((1-\kappa_{S\Phi})^2-\kappa_{S}(1-2\kappa_{S\Phi}))}{9\,(4/27 - \xi^2)}
\left( \frac{M_{\xi}}{m_P} \right)^2 + \cdots \right) H^2,
\end{equation}
which can spoil inflation for $\kappa_S \sim 1$, but
for reasonably natural values of parameters $\kappa_S \lesssim 0.01$ and
$|\kappa_{S\Phi}| \lesssim 1$ we can obtain successful inflation
consistent with WMAP7 data.

\begin{figure}[th]
\begin{center}
\epsfig{file=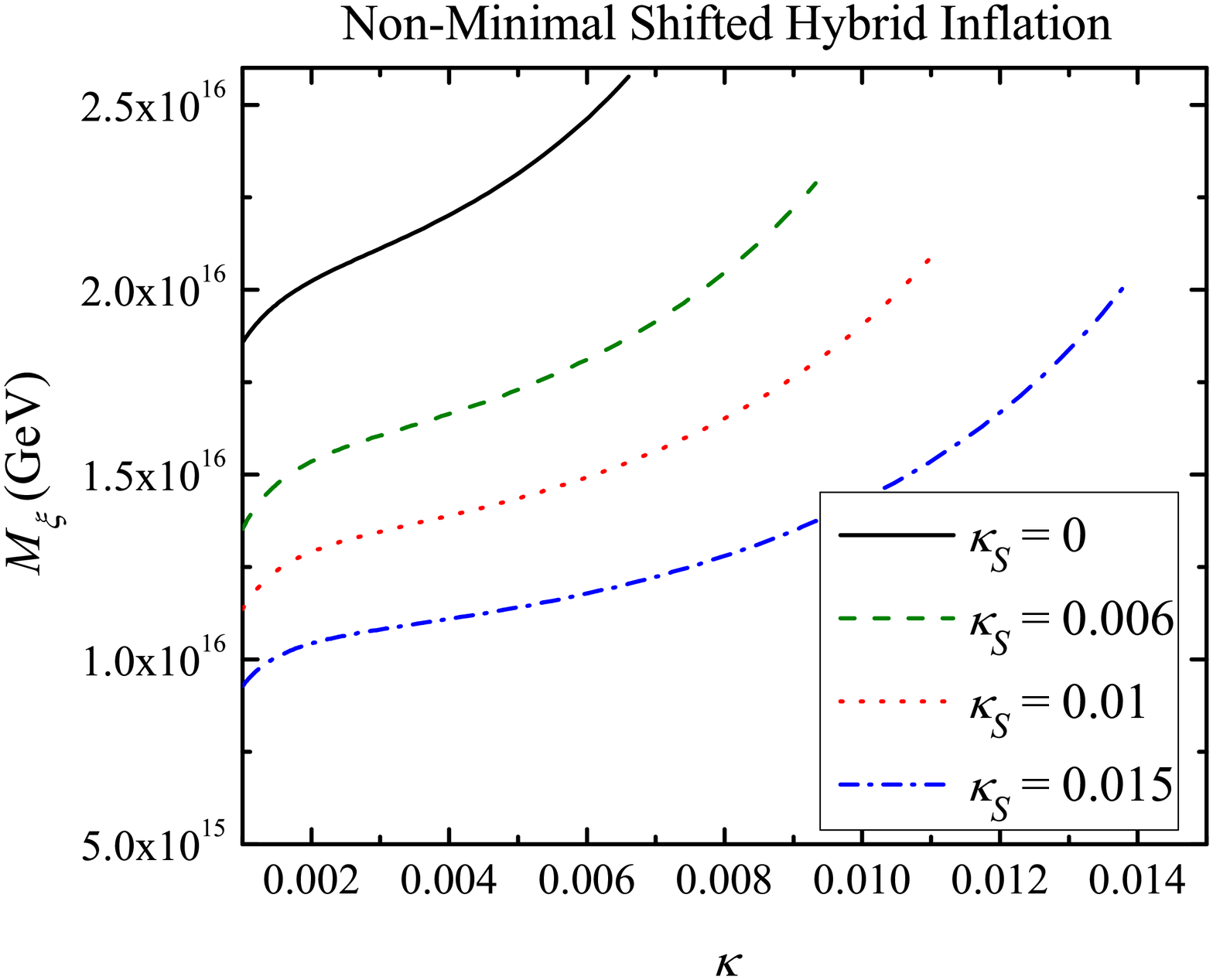, width=8cm}
\epsfig{file=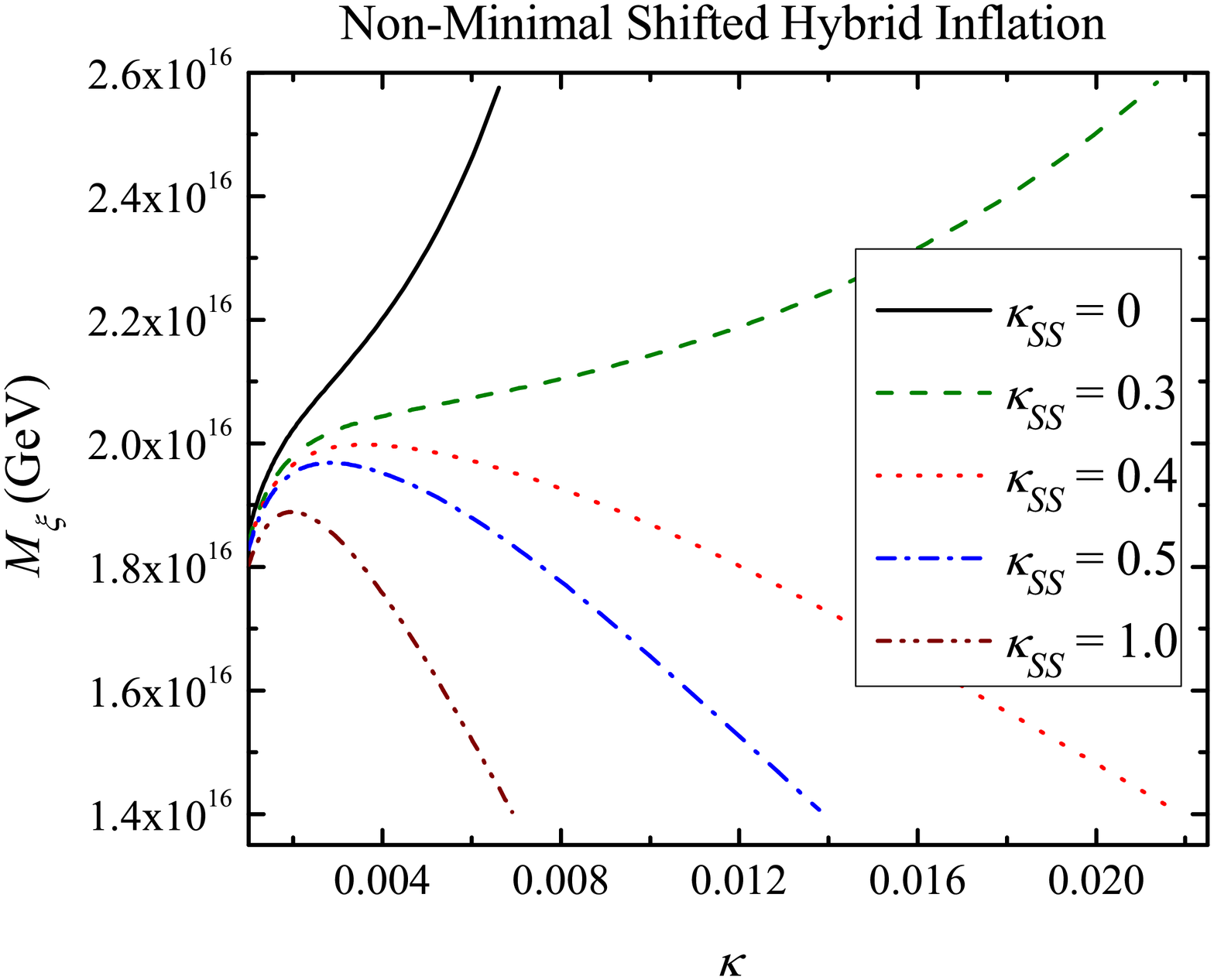, width=8cm}
\end{center}
\caption{$M_{\xi}$ vs $\kappa$
for shifted hybrid inflation with $\xi = 0.3$ and $T_r = 10^{9}$ GeV.}
\label{fig3}
\end{figure}

\begin{figure}[th]
\begin{center}
\epsfig{file=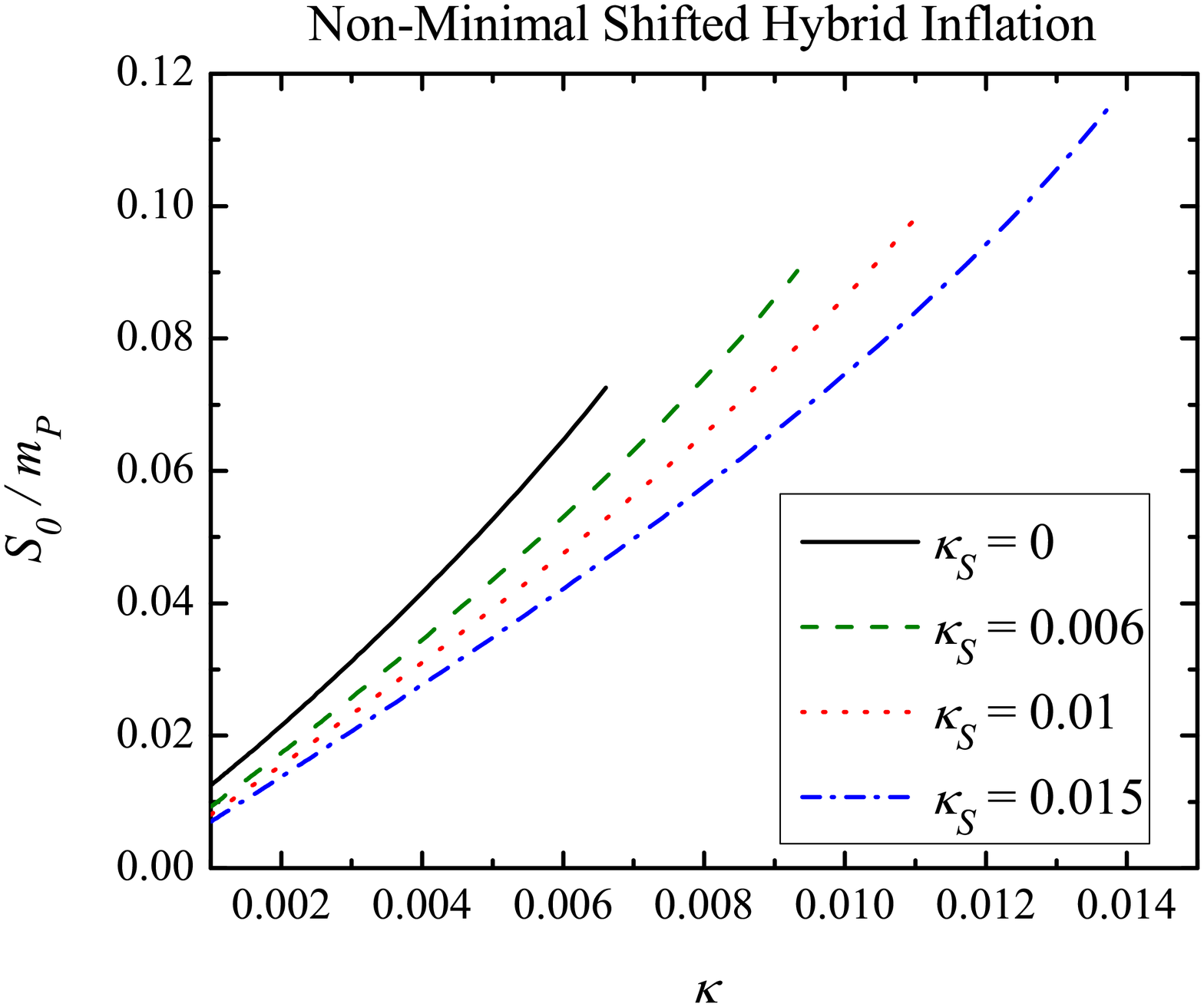, width=8cm}
\epsfig{file=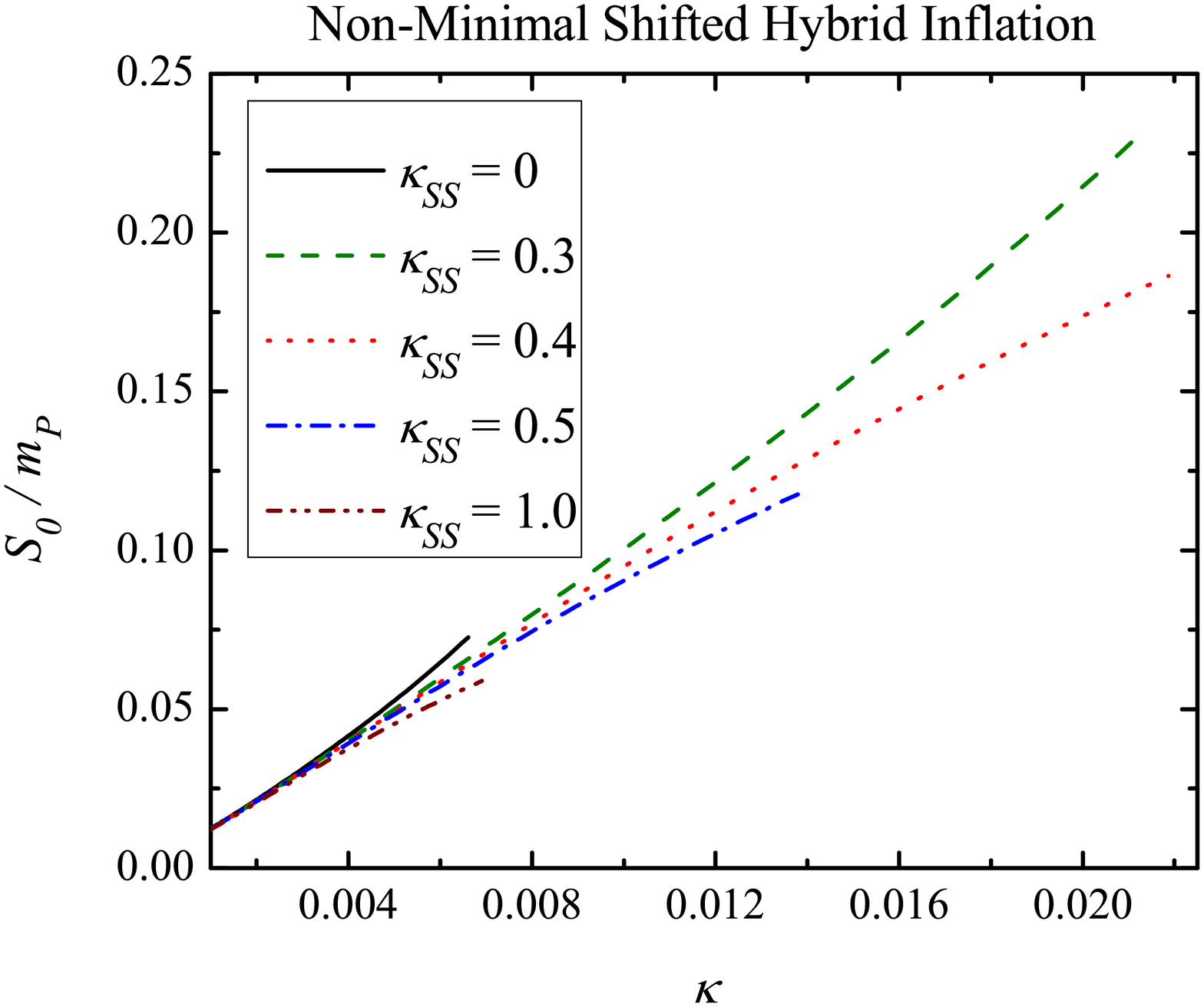, width=8cm}
\end{center}
\caption{$S_0/m_P$ vs $\kappa$
for shifted hybrid inflation with $\xi = 0.3$ and $T_r = 10^{9}$ GeV.}
\label{fig4}
\end{figure}

The predicted values of various parameters for shifted hybrid 
inflation are displayed in Figs.~\ref{fig2}, \ref{fig3}
and \ref{fig4}. For minimal K\"ahler potential the scalar 
spectral index $n_s \gtrsim 0.99$ lies outside the 1-$\sigma$ 
bounds of WMAP7 data. With non-minimal K\"ahler potential
only $\kappa_S$ and $\kappa_{SS}$ play the important role
of reducing the scalar spectral index to the central value
of WMAP7 data, i.e. $n_s \simeq 0.96$. As can be seen in 
Fig.~\ref{fig2} we can obtain $n_s$ within the 1-$\sigma$ 
bounds of WMAP7 data for $\kappa_S \gtrsim 0.005$ or
$0.33 \lesssim \kappa_{SS} \lesssim 1$ with 
all other non-minimal parameters equal to zero.

The $\kappa_S \neq 0$ case has been considered
previously in Ref.~\cite{urRehman:2006hu} 
for standard and smooth hybrid inflation. The results
we obtain here are quite similar to the one obtained 
in Ref.~\cite{urRehman:2006hu}. In particular, the scalar
spectral index is given by
\be
n_s \simeq 1 - 2\,\kappa_S + 
\left( \frac{8(1-\kappa_{S})}{9\,(4/27 - \xi^2)}
+6\gamma_S x_0^2 \right)
\left( \frac{M_{\xi}}{m_P} \right)^2
- \frac{275 \kappa^2}{16\,\pi^2}
\left| \partial^2_{x_0} F(5\,x_0^2) \right|
 \left( \frac{m_P}{M_{\xi}} \right)^2.
\ee
From Fig.~\ref{fig3} and Fig.~\ref{fig4} it is clear that 
the value of parameters $|S_0|/m_P$, $M_{\xi}/m_P$ and 
$x_0 \left(= |S_0|/M_{\xi}\right)$ increases with $\kappa$. 
Therefore, the sugra contribution in the above expression raises
the value of the scalar spectral index $n_s$ with $\kappa$. 
The radiative correction, on the other hand, does not vary 
much with $\kappa$ since both $|F''|$ ($\simeq \frac{1}{x_0^2}$ 
for large $x_0$) and $\left( \frac{m_P}{M_{\xi}} \right)^2$ 
tries to compensate the increase 
in $\kappa^2$ term, in above expression. For small
values of $\kappa$ sugra correction is suppressed
as compared to radiative correction and $-2\,\kappa_S$
factor, which reduces the scalar spectral index
within the 1-$\sigma$ bounds of WMAP7 data.

For $\kappa_{SS} \neq 0$ case we obtain 
following result for the scalar spectral index:
\be
n_s \simeq 1 + 
\left( \frac{8}{9\,(4/27 - \xi^2)}
 + 6\,(1 - 3\,\kappa_{SS})\,x_0^2 \right)
\left( \frac{M_{\xi}}{m_P} \right)^2
- \frac{275 \kappa^2}{16\,\pi^2}
\left| \partial^2_{x_0} F(5\,x_0^2) \right|
 \left( \frac{m_P}{M_{\xi}} \right)^2.
\ee
In this case with $\kappa_{SS} \lesssim 1/3 \simeq 0.33$  
sugra term is positive and raises the value of $n_s$ with
$\kappa$. For small values of $\kappa$ (or $x_0$)
radiative correction becomes important and a small
reduction in $n_s$ is observed as shown in Fig.~\ref{fig2}. 
On the other hand, for $\kappa_{SS} \gtrsim 0.33$,
$\gamma_S$ becomes negative and we obtain a reduction in 
$n_s$ with $\kappa$ which is consistent with the WMAP7 data 
(see Fig.~\ref{fig3}). With $\kappa_{SS} \gtrsim 0.33$
quadratic and quartic terms of the potential in Eq.~(\ref{VT}) 
are positive and negative respectively. This provides a nice 
example of hilltop inflation \cite{Boubekeur:2005zm} and a 
similar kind of potential has been analyzed in 
Refs.~\cite{Kohri:2007gq,Lin:2009yt,Pallis:2009pq,Rehman:2009wv}.
The value of the tensor to scalar ratio $r \lesssim 10^{-5}$
remains small in the above model of shifted hybrid inflation.
For realizing observable $r$ values in supersymmetric hybrid
inflationary models see Ref.~\cite{Shafi:2010jr}.

\section{\large{\bf Gauge coupling unification 
and TeV scale vector-like matter}}
In this section we discuss the impact of the massless 
$SU(3)$ octet and $SU(2)$ triplet multiplets on the 
gauge coupling unification. 
After $SU(5)$ breaking, these multiplets
remain massless in the limit of exact supersymmtery 
and may spoil gauge coupling unification.
After inclusion of soft SUSY breaking mass terms taking account of 
$\langle S \rangle \sim $ TeV, these particles acquire masses
of order TeV.

In order to achieve gauge coupling unification we can add
suitable vectorlike particles with TeV scale masses.
These vectorlike particles have recently been studied to 
solve the little hierarchy problem in the MSSM 
\cite{Babu:2008ge,Martin:2009bg,Graham:2009gy}.
The requirement that the three gauge couplings should 
remain perturbative at least up to the unification scale 
and the value of strong coupling should lie 
within the experimental uncertainties \cite{pgd},
greatly reduces the choices of vectorlike combinations.
Furthermore, in order to avoid fast proton decay we do 
not consider the triplet scalar Higgs vectorlike combination 
$D(3, 1, -1/3) + D(\overline{3}, 1, 1/3)$. Taking into account
these considerations we choose the following combination of
extra vectorlike particles:
\be \label{comb}
L(1,2,1/2)+\overline{L}(1,2,-1/2)
+ 2(E(1,1,1)+\overline{E}(1,1,-1)). \ee 
The sum of 1-loop beta function of octet $G(8,1,0)$, 
triplet $W(1,3,0)$ and the above vectorlike combination is 
$\Delta b = (0,0,3)+(0,2,0)+(3,1,0)=(3,3,3)$. The 2-loop 
beta functions and RGEs for SM, MSSM and these extra 
vectorlike particles can be found in 
Refs.~\cite{mac,Cvetic:1998uw,Barger:1992ac,Martin:1993zk,Barger:2007qb}.

\begin{figure}[t]
\centering \includegraphics[width=8cm]{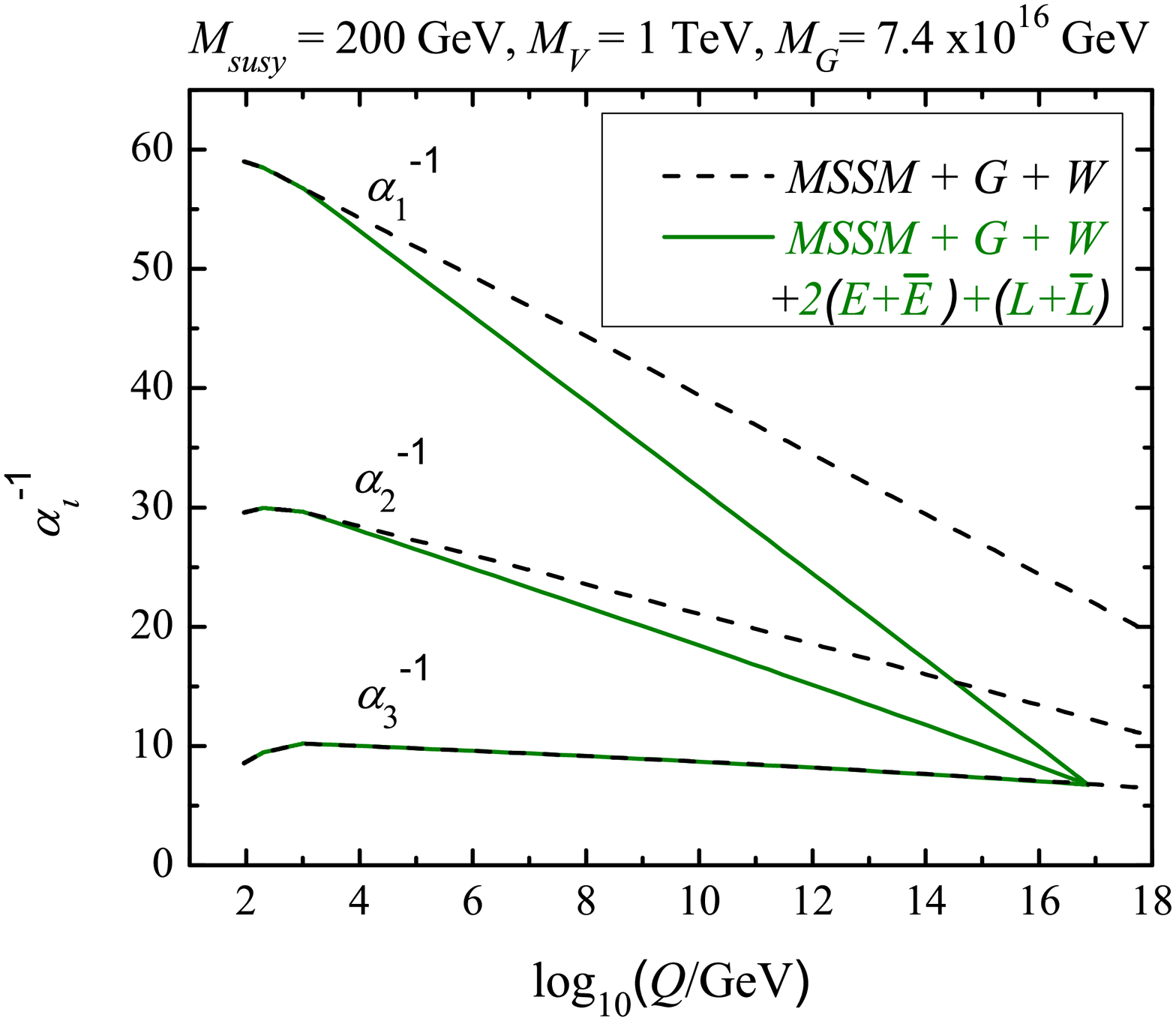}
\centering \includegraphics[width=8cm]{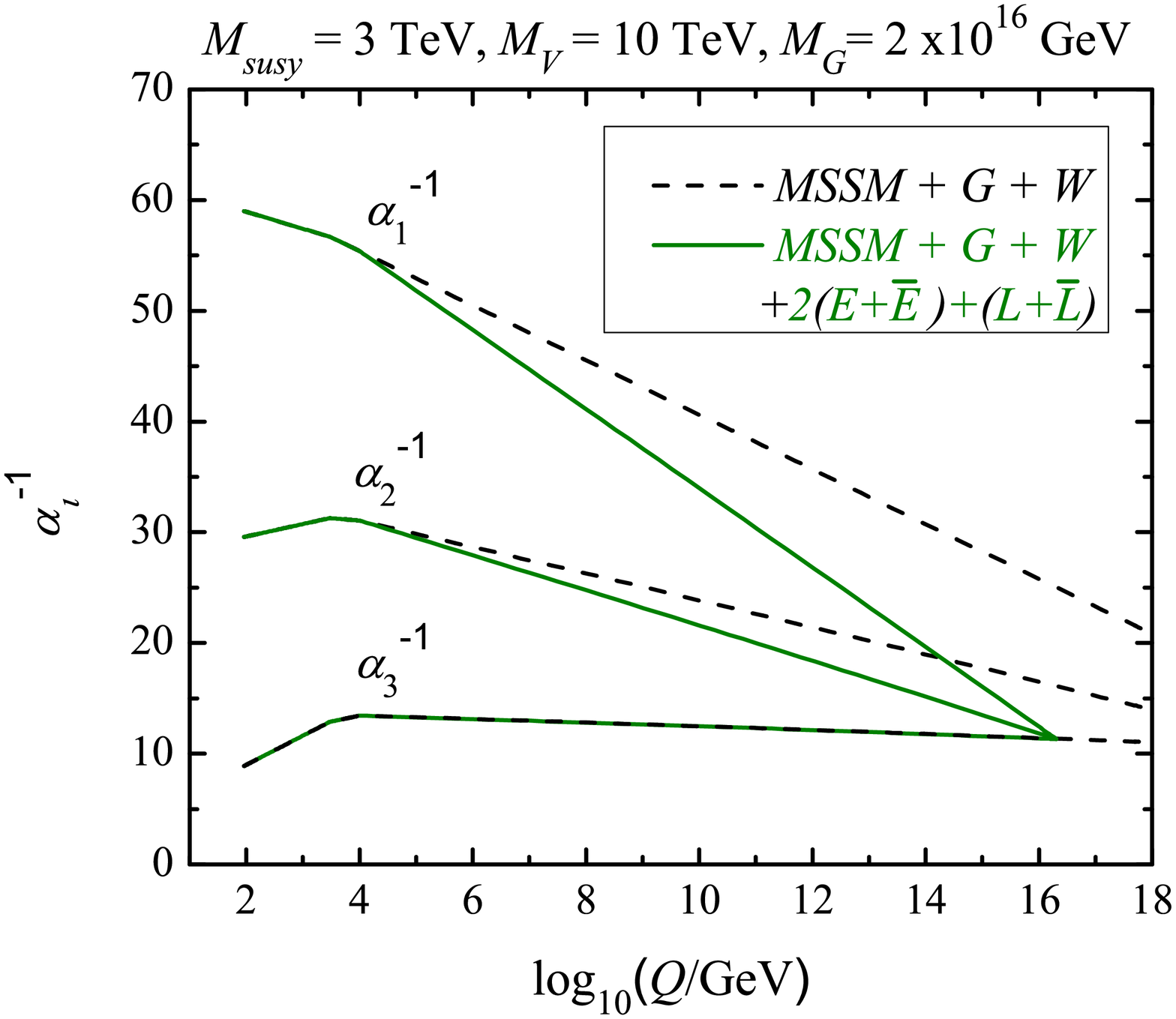}
\caption {Gauge coupling evolution with the 
effective SUSY breaking scale $M_{susy} = 200$ GeV (left panels), 
$M_{susy} = 3$ TeV (right panels) and $\tan \beta = 10$.
Dotted (solid) lines correspond to $MSSM+G+W$
($MSSM+G+W+(L+\overline{L})+2(E+\overline{E})$).
The masses of $G$, $W$ and extra vectorlike particles 
are set equal to $M_V = 1$ TeV (left panels) and 
$M_V = 10$ TeV (right panels).}
\label{fig7}%
\end{figure}

The evolution of three gauge couplings with and without the extra
vectorlike particle combination (Eq.~(\ref{comb})) is shown 
in Fig.~\ref{fig7}. Here we have used 2-loop RGEs with 
the SUSY breaking scale $M_{susy} = 200$ GeV and $M_{susy} = 3$ TeV
and the masses of vectorlike particles $M_V = 1$ TeV and
$M_V = 10$ TeV. The value of the strong gauge coupling is fixed
by the gauge unification condition and is required to lie
within the experimental uncertainties \cite{pgd}. The GUT
scale is found to lie in the range 
$M_G \sim (2-7)\times 10^{16}$ GeV
for $M_{susy} \sim (0.2-3)$ TeV and $M_V \sim (1-10)$ TeV.
These extra particles may be detected at the LHC
provided their masses are less than or of order a TeV 
\cite{Han:2010rf}.
As an example, the gluon-gluon fusion channel can
lead to $SU(3)$ octet pair production at the LHC:
\begin{center}
$A^{\mu}_i \, A^{\mu}_i\, \longrightarrow \,\phi_i \, \phi_i\, 
\longrightarrow  \,q \, \bar{q}\,\,\, (e^{+} \, e^{-}),$
\end{center}
where $i = 1,...,8$. This coupling can be generated from 
the kinetic energy term of $\Phi$.

\section{\large{\bf Summary}}
To summarize, we have analyzed the adjoint field hybrid 
inflationary model in supersymmetric $SU(5)$. 
Since the minimal 
SUSY $SU(5)$ hybrid inflation suffers from 
the monopole problem we have discussed $SU(5)$ 
shifted hybrid inflation.
In this model the $SU(5)$ gauge symmetry is 
broken during inflation and monopoles are inflated 
away before inflation ends.
In minimal shifted hybrid inflation the spectral 
index $n_s \gtrsim 0.99$ lies outside the WMAP7 
1-$\sigma$ bounds with $\kappa \gtrsim 10^{-3}$ and 
TeV scale soft SUSY breaking masses. 
However, with a nonminimal K\"ahler potential 
the WMAP7 1-$\sigma$ region 
nicely compatible with shifted hybrid inflation.
A tiny value of $r \lesssim 10^{-5}$ is obtained
in this model\footnote{ For a discussion of
observable $r$ in these supersymmetric models see 
Ref.~\cite{Shafi:2010jr}}. 
Furthermore, as a consequence of $R$-symmetry, the 
$SU(2)$ octet and $SU(3)$ triplet supermultiplets
lie in the TeV range. In order to preserve 
gauge coupling unification we include additional
TeV scale vectorlike particles which may be 
observed at the LHC.

\section*{Acknowledgements}%

We thank Joshua R. Wickman for valuable discussions.
The work of S. K. was partially supported by the 
Science and Technology Development Fund (STDF) Project 
ID 437 and the ICTP Project ID 30. 
Q.S and M. R acknowledge partial support from an NSF 
collaborative grant with Egypt, award number 0809789,
and support from the DOE under grant 
No.~DE-FG02-91ER40626 (Q.S. and M.R.), and by the
University of Delaware (M.R.).



\end{document}